\documentstyle[11pt]{article}
\include{epsf}
\include{rotate}
\newcommand{\bc}{\begin{center}}
\newcommand{\ec}{\end{center}} 
\newcommand{\bt}{\begin{tabular}}
\newcommand{\et}{\end{tabular}}
\newcommand{\bdes}{\begin{description}}
\newcommand{\edes}{\end{description}}

\newcommand{\ul}{\vec}
\newcommand{\be}{\begin{equation}}
\newcommand{\ee}{\end{equation}}
\newcommand{\bea}{\begin{eqnarray}}
\newcommand{\eea}{\end{eqnarray}}

\newcommand{\etal}{{\it et\ al.\ }}

\newcommand{\ba}{\begin{array}}
\newcommand{\ea}{\end{array}}
\newcommand{\SG}{Si$_x$Ge$_{1-x}$ }
\newcommand{\hhp}{\vert{\frac 3 2}, +{\frac 3 2}>}
\newcommand{\hhm}{\vert{\frac 3 2}, -{\frac 3 2}>}
\newcommand{\lhp}{\vert{\frac 3 2}, +{\frac 1 2}>}
\newcommand{\lhm}{\vert{\frac 3 2}, -{\frac 1 2}>}
\title{FINITE ELEMENT ANALYSIS OF STRAIN EFFECTS ON ELECTRONIC
AND TRANSPORT PROPERTIES IN QUANTUM DOTS AND WIRES}
\author{ {H. T. Johnson, L. B. Freund, C. D. Aky\"{u}z, and A. Zaslavsky} \\  Division of Engineering, Brown University, Providence, RI 02912} 
\date{}
\begin{document}
\setcounter{page}{1}
\setlength{\baselineskip}{14pt}
\maketitle
\normalsize
\baselineskip=14pt
\begin{abstract}

\baselineskip=14pt 
Lattice mismatch in epitaxial layered heterostructures with small
characteristic lengths induces large, spatially nonuniform strains.
The components of the strain tensor have been shown experimentally to
affect the electronic properties of semiconductor structures.  Here a
technique is presented for calculating the influence of strain on
electronic properties.  First the linear elastic strain in a quantum
dot or wire is determined by a finite element calculation.  A
strain-induced potential field that shifts and couples the valence
subbands in the structure is then determined from deformation
potential theory.  The time-independent Schr\"{o}dinger equation,
including the nonuniform strain induced potential and a potential due
to the heterostructure layers, is then solved, also by means of the
finite element method.  The solution consists of the wavefunctions and
energies of states confined to the active region of the structure;
these are the features which govern the electronic and transport
properties of devices.  As examples, two \SG submicron resonant
tunneling devices, a quantum wire with two dimensional confinement and
a quantum dot with three dimensional confinement, are analyzed.
Experimentally measured resonant tunneling current peaks corresponding
to the valence subbands in the material are modeled by generating
densities of confined states in the structures.  Size and composition
dependent strain effects are examined for both devices.  In both the
quantum dot and the quantum wire, the strain effects on the
wavefunctions and energies of confined states are evident in the
calculated densities of confined states in the structures, which are
found to be consistent with experimentally measured tunneling
current/voltage curves for resonant tunneling diodes.

\end{abstract}

\newpage

\section*{I.  INTRODUCTION}

Epitaxially grown semiconductor heterostructures often consist of
materials with lattice parameters that are mismatched by as much as
several percent.  For thin films of large lateral extent, these
strains are spatially uniform and the effects are well understood.
However, structures of relatively small lateral extent, having
distinctive geometric features and bounded by free surfaces, are
characterized by strains that are highly nonuniform.  The effects of
nonuniform strain on the electronic properties of semiconductor
heterostructures have been observed experimentally, but the coupled
physical phenomenon has not been extensively modeled.$^{1-4}$ The
analysis of strain effects in a quantum mechanical model of
semiconductor devices has only recently been attempted by Pryor
\etal$^{5-7}$ and Williamson \etal$^{8}$, who calculate strain induced
potentials and wavefunctions in quantum dots, and Zunger$^{9}$, who
reviews the topic of electronic structure in pyramidal semiconductor
quantum dots based on atomistic methods.

Strained semiconductor devices that are based on quantum effects,
particularly charge confinement in one or more spatial dimensions,
underlie a potentially significant technology.  Much can be learned
about quantum effects by studying the class of devices based on
resonant tunneling of carriers from an emitter region, into a quantum
well, and then into a collector region.  The material combinations in
these devices and their geometrical features, including the layered
structure and the free surfaces, lead to complicated mechanical strain
fields.  Because the sequential tunneling of carriers is a simple
phenomenon governed by the spectrum of available states in the quantum
well, and because the devices are extremely small and the operating
temperatures are low, it is likely that the effects of strain on the
electronic and transport properties can be represented quantitatively
through modeling.

Calculations of elastic strain fields in semiconductor structures are
well suited for the finite element method, which is a common tool in
continuum mechanics.$^{2,10}$ In the technique presented here, the
finite element calculation of the strain in a device is made using a
general purpose structural mechanics finite element package.$^{11}$
However, the use of the finite element method (FEM) in quantum
mechanics, which is reviewed by Linderberg $^{12}$, is much less common.
Models of semiconductor devices by means of FEM have been proposed by
a number of authors for one and two dimensional problems.  Several FEM
models are available for one dimensional resonant tunneling structures
which include the effects of arbitrary potential profiles due to
layered composition.$^{13-15}$ Chen$^{16}$ models a one dimensional
resonant tunneling diode using FEM and calculates a current-voltage
curve based on a quantum hydrodynamic model.  Electron wavefunctions
and band structures for two dimensional quantum wires or quantum dots
are analyzed using FEM by a number of investigators$^{17-23}$.
Tsuchida and Tsukada$^{24}$ calculate the electronic structure of a
perfect Si lattice using a three dimensional FEM formulation.
However, none of the studies consider a three dimensional device, and
none have considered the effects of strain on experimentally measured
electronic properties of devices.

The finite element method is well suited for finding approximate
solutions of boundary value problems for partial differential
equations in finite domains, especially if the (unknown) exact
solution is a minimizer of a total energy functional.  Both the stress
boundary value problem and the quantum mechanical boundary value
problem in the present study are of this type.  The central idea of
the method is that an unknown continuous field in the domain is
represented approximately in terms of its values at discrete points
(nodes) within the domain; the goal is to determine optimal values for
these nodal quantities.  The domain is covered with areas (in two
dimensions) or volumes (in three dimensions) whose boundaries are
defined geometrically by the nodes; these areas or volumes are the
elements.  Fields are defined within each element in terms of the
values of the nodal quantities on its boundary by means of a suitable
interpolation scheme.  With the complete field defined in terms of the
nodal quantities, the total energy can be expressed in terms of the
global vector of nodal quantities.  Imposition of the condition that
the actual values of the nodal quantities must render the total energy
a minimum leads to a system of algebraic equations for these values.
The method is ideally suited for numerical analysis by computer.  In
general, it is convergent as nodal spacing diminishes for elliptic
partial differential equations which have unique solutions.  It should
be noted that the method is of far broader applicability than is
implied by these introductory comments.

In this work, a finite element model is used to analyze both the
continuum mechanics and the quantum mechanics of a strained
semiconductor device.  The strain field is shown to affect the
performance of the device.  The model is used to analyze devices
studied experimentally by Aky\"{u}z {\etal}$^2$ and Lukey
{\etal}$^{4}$, who show the effects of nonuniform strain on the
resonant tunneling current vs. voltage characteristics of \SG quantum
wires and quantum dots, shown schematically in Figure~1.  The main
features of the model are described in Section II; these include the
strain calculation, the treatment of the strain effects by deformation
potential theory, and the quantum mechanical model.  The finite
element formulation for the two or three dimensional Schr\"{o}dinger
equation incorporating valence subband coupling and the nonuniform
potential is outlined in Section III.  The results of the calculations
for the resonant tunneling structures are discussed in Section IV, and
comparisons are made between the calculations and the available
experimental results.

\section*{II.  CONTINUUM AND QUANTUM MECHANICAL MODELS}

The analysis of a strained semiconductor heterostructure is divided
into three calculations.  First, a linear elastic finite element
calculation is made to determine the strain field, which is a function
of the composition and the geometry of the structure.  Second, the
strain induced potential field is calculated using deformation
potential theory.  Third, the time-independent Schr\"{o}dinger
equation including the strain induced potential is solved numerically
by means of the finite element method to obtain the spectrum of
energies and wavefunctions of available states.

\subsection*{A.  Strain Field Calculation}

The strain field due to the constraint of epitaxy associated with the
mismatched lattice parameters of the heterostructure layers is
determined within the framework of linear elasticity theory.  The
structure is discretized spatially with a mesh which is more refined
near free surfaces and in regions where the mismatch between adjacent
layers is larger.  The finite element mesh used for a strain
calculation in the resonant tunneling diode quantum wire is shown in
Figure~2.  The mismatch condition is imposed in the finite element
calculation by prescribing in each layer a uniform stress-free
dilatation that is proportional to the bulk lattice parameter of the
material in that layer.  Continuity of displacements is required
across the layer interfaces; it is this constraint which gives rise to
stress.  The outer surfaces are considered to be traction free,
implying that certain components of the stress tensor vanish, and the
material is allowed to relax until it reaches an equilibrium
configuration.  Thus, using a standard structural mechanics finite
element program $^{11}$, the complete state of stress, strain and
displacement is determined throughout the device.  Strain components
are shown in Figures 3 and 4 for the resonant tunneling diode quantum
wire.  Strain is a tensor quantity, and its components show
significant variation with position throughout the device.  In smaller
devices, the strain is more nonuniform due to the pervasive effect of
relaxation at the free surfaces.

\subsection*{B.  Strain Induced Potential}

The components of strain induce a potential field that affects the
wavefunctions and energies of the charge carriers in an otherwise
perfect crystal.  From first order perturbation theory, the strain
induced potential that affects wavefunctions in subbands $\alpha$ and
$\beta$ is formed from the tensor product (Singh$^{25}$) 
\be
V_\epsilon^{\alpha\beta}(\ul{r})={D_{ij}}^{\alpha\beta}\epsilon_{ij}(\ul{r})
\label{strainpot}
\ee where $\ul{r}$ is an arbitrary position vector,
$\epsilon_{ij}(\ul{r})$ is the strain tensor field, and
$D_{ij}^{\alpha\beta}$ is the deformation potential tensor for
subbands $\alpha$ and $\beta$, which consists of components derived
experimentally.  The indices $ij$ range over the coordinate
directions.  For the \SG material combination, the valence band
electronic properties are dominated by the heavy hole and light hole
subbands, so the $\alpha\beta$ basis includes the heavy hole (HH)
subbands denoted by $\mid{\frac 3 2}, \pm {\frac 3 2}>$, and the light
hole (LH) subbands denoted by $\mid{\frac 3 2}, \pm {\frac 1 2}>$.
The split-off subbands (SO) are ignored because the separation energy is
considered to be large enough so that coupling effects can be neglected.  
The potential equation can be written as

\be
V_\epsilon^{\alpha\beta}(\ul{r})=
\bordermatrix{& {\hhp} & {\hhm} & {\lhp} & {\lhm} \cr
{\hhp} & D_{ij}^{11}(\ul{r})\epsilon_{ij}(\ul{r}) & D_{ij}^{12}(\ul{r})\epsilon_{ij}(\ul{r}) & D_{ij}^{13}(\ul{r})\epsilon_{ij}(\ul{r}) & D_{ij}^{14}(\ul{r})\epsilon_{ij}(\ul{r}) \cr
{\hhm} & D_{ij}^{21}(\ul{r})\epsilon_{ij}(\ul{r}) & D_{ij}^{22}(\ul{r})\epsilon_{ij}(\ul{r}) & D_{ij}^{23}(\ul{r})\epsilon_{ij}(\ul{r}) & D_{ij}^{24}(\ul{r})\epsilon_{ij}(\ul{r}) \cr
{\lhp} & D_{ij}^{31}(\ul{r})\epsilon_{ij}(\ul{r}) & D_{ij}^{32}(\ul{r})\epsilon_{ij}(\ul{r}) & D_{ij}^{33}(\ul{r})\epsilon_{ij}(\ul{r}) & D_{ij}^{34}(\ul{r})\epsilon_{ij}(\ul{r}) \cr
{\lhm} & D_{ij}^{41}(\ul{r})\epsilon_{ij}(\ul{r}) & D_{ij}^{42}(\ul{r})\epsilon_{ij}(\ul{r}) & D_{ij}^{43}(\ul{r})\epsilon_{ij}(\ul{r}) & D_{ij}^{44}(\ul{r})\epsilon_{ij}(\ul{r}) \cr
} 
\label{strpot}
\ee
 
\noindent
where the terms $D_{ij}^{\alpha \beta}$ are the deformation potential
tensors which range over the spatial dimensions $i$ and $j$.  The
deformation potential tensors contain material constants which vary
spatially as the composition varies in the device; values of these
constants for a wide range of materials are known from experiments.
The repeated $ij$ indices indicate the scalar product (contraction
over the product of second rank tensors) between the deformation
potential tensor and the strain tensor.  Details of the
$D_{ij}^{\alpha \beta}$ terms are given in Appendix A, along with the
material constants used in the calculations.  Thus, for a calculated
strain tensor function of position $\epsilon_{ij}(\ul{r})$, it is
possible to calculate the deformation potential function of position
$V_\epsilon^{\alpha\beta}(\ul{r})$ to be included in the quantum
mechanical analysis.

\subsection*{C.  Quantum Mechanical Model}

In resonant tunneling structures, tunneling currents are determined by
available quantized states for individual charge carriers.  The
energies and wavefunctions of a single carrier in the semiconductor
structure are solutions of the time independent Schr\"{o}dinger
equation \be H_{k \cdot p}^{\alpha
\beta}(\ul{r})\Psi^{\beta}(\ul{r})+V^{\alpha
\beta}(\ul{r})\Psi^{\beta}(\ul{r})=E\Psi^{\alpha}(\ul{r})
\label{sch}
\ee 
where $\Psi^{\alpha}(\ul{r})$ is the wavefunction in subband
$\alpha$, $E$ is the energy, $H_{k \cdot p}^{\alpha \beta}(\ul{r})$ is
the {\bf k} $\cdot$ {\bf p} Hamiltonian operator, and $V^{\alpha
\beta}(\ul{r})$ is a potential function of position.

The {\bf k} $\cdot$ {\bf p} perturbation method is used to model the
medium.  Using this technique, there is a tensor function
for the effective mass associated with each subband, and there are
{\bf k} $\cdot$ {\bf p} terms coupling the effective masses in
different subbands.  Written in the same form as (\ref{strpot}) the
Hamiltonian is

\be
H_{k \cdot p}^{\alpha\beta}(\ul{r})=
\bordermatrix{& {\hhp} & {\hhm} & {\lhp} & {\lhm} \cr
{\hhp} & L_{ij}^{11}(\ul{r})\nabla^2_{ij} & L_{ij}^{12}(\ul{r})\nabla^2_{ij} & L_{ij}^{13}(\ul{r})\nabla^2_{ij} & L_{ij}^{14}(\ul{r})\nabla^2_{ij} \cr
{\hhm} & L_{ij}^{21}(\ul{r})\nabla^2_{ij} & L_{ij}^{22}(\ul{r})\nabla^2_{ij} & L_{ij}^{23}(\ul{r})\nabla^2_{ij} & L_{ij}^{24}(\ul{r})\nabla^2_{ij} \cr
{\lhp} & L_{ij}^{31}(\ul{r})\nabla^2_{ij} & L_{ij}^{32}(\ul{r})\nabla^2_{ij} & L_{ij}^{33}(\ul{r})\nabla^2_{ij} & L_{ij}^{34}(\ul{r})\nabla^2_{ij} \cr
{\lhm} & L_{ij}^{41}(\ul{r})\nabla^2_{ij} & L_{ij}^{42}(\ul{r})\nabla^2_{ij} & L_{ij}^{43}(\ul{r})\nabla^2_{ij} & L_{ij}^{44}(\ul{r})\nabla^2_{ij} \cr
}
\label{effmass}
\ee

\noindent
where the $L_{ij}^{\alpha \beta}$ tensors on the diagonal of the
matrix are the effective mass tensors for each subband, and the off
diagonal $L_{ij}^{\alpha \beta}$ tensors introduce a {\bf k} $\cdot$
{\bf p} coupling of subbands.  Like the terms in the deformation
potential tensors, the components of $L_{ij}^{\alpha \beta}$ contain
material constants and thus vary spatially throughout the device.  The
exact forms of the $L_{ij}^{\alpha \beta}$ tensors, which come from
the Luttinger-Kohn Hamiltonian, are given in Appendix A.

The deformation potential and effective mass material constants in
(\ref{strpot}) and (\ref{effmass}) are functions of the local
composition of the device.  Values for the constants in each layer are
given by linear interpolation of the material constants associated
with each of the pure elements composing that layer.

The nonuniform potential $V^{\alpha \beta}(\ul{r})$ consists of
contributions from two sources and is given by \be V^{\alpha
\beta}(\ul{r})=V_{c}^{\alpha \beta}(\ul{r})+V_\epsilon^{\alpha
\beta}(\ul{r}) \ee where $V_{c}^{\alpha \beta}(\ul{r})$ is due to the
valence band alignment of material at a given position in the device, and
$V_\epsilon^{\alpha \beta}(\ul{r})$ is the strain induced potential
given in (\ref{strainpot}).  The strain induced potential
$V_\epsilon^{\alpha \beta}(\ul{r})$, like the components of the strain
tensor, is in general nonuniform in both the lateral directions and
the vertical direction in the structure.  The composition based band
offset potential $V_{c}^{\alpha \beta}(\ul{r})$ is nonuniform only in
the growth, or vertical direction in the structure.  The total
potential $V^{\alpha \beta}(\ul{r})$ is shown in Figure~5 for a
representative quantum dot calculation.  

\section*{III.  FINITE ELEMENT TECHNIQUE FOR THE SCHR\"{O}DINGER EQUATION}

\subsection*{A.  Finite Element Formulation}

The physical domain of the device is discretized into a mesh of nodes
and elements.  Elements used for the quantum wires and quantum dots
are described in Appendix B.  The mesh extends to the free or
insulated surfaces, which impose an infinite potential on the
wavefunction.  The wavefunctions and energies of the states localized
in the active region are insensitive to remote boundary conditions,
{\it i.e.} conditions at boundaries located a large distance away
relative to the active region size.  The mesh is more refined in the
active region of the device, where large wavefunction gradients are
expected.

The form of the Schr\"{o}dinger equation to be solved on the finite
element mesh is obtained by minimizing the total variation of the weak
form of the equation with respect to the wavefunction.  The minimum in
variation with respect to the wavefunction corresponds physically to a
minimum energy.  Details of the complete variational formulation of
the finite element technique are included in Appendix B.  The
functional corresponding to the weak form of the time independent
Schr\"{o}dinger equation with a nonuniform potential (\ref{sch}) is
given by \be \Pi(\Psi^{\alpha})=\int_R \nabla \Psi^{\alpha}
L^{\alpha\beta} \nabla \Psi^{\beta} dR + \int_R
\Psi^{\alpha}V^{\alpha\beta}\Psi^{\beta} dR - E\int_R
\Psi^{\alpha}\Psi^{\beta} dR
\label{var}
\ee where $\Psi^{\alpha}$, $L^{\alpha\beta}$, and $V^{\alpha\beta}$
are functions of position in the structure.  The term
$L^{\alpha\beta}$ is taken to be constant within each element of
the mesh.  The fields $\Psi^{\alpha}$, $\nabla \Psi^{\alpha}$, and
$V^{\alpha\beta}$ are represented by their nodal values. Values
throughout each element are determined by interpolation according to
the particular shape functions that are adopted.  Thus, for shape
functions $N(\ul{r})$ used here, these fields are written as 
\bea
\Psi^{\alpha}=\sum_{A=1}^{all\;nodes}\Psi_A^{\alpha}N_A(\ul{r})\;\;\;\;\;\;\;\nabla\Psi^{\alpha}=\sum_{B=1}^{all\;nodes}\Psi_B^{\alpha}\nabla
N_B(\ul{r})\;\;\;\;\;\;\;
V^{\alpha\beta}=\sum_{C=1}^{all\;nodes}V_C^{\alpha\beta}N_C(\ul{r})
\label{disc}
\eea where the coefficients in the summations are the values of the
fields at the individual nodes.  The functional $\Pi(\Psi^{\alpha})$
is rewritten in a discrete form using (\ref{disc}) and then minimized
with respect to the nodal values of the wavefunction,
$\Psi_A^{\alpha}$.
Integrals over the region $R$ of the entire structure are replaced
with integrals over individual element volumes ($\Omega^e$) and a
summation over all elements, so that the finite element form of
the Schr\"{o}dinger equation becomes
\be
\sum^{elems} \left [ \sum_A \sum_B \int_{\Omega^e} \left(\nabla
N_A L^{\alpha\beta}\nabla N_B + N_A N_B \sum_C V_C^{\alpha\beta} N_C \right)d\Omega
\right ] \Psi_B^{\beta} = E \sum^{elems} \left [
\sum_A \sum_B \int_{\Omega^e}N_A N_B d\Omega \right ] \Psi_B^{\alpha}
\label{femsch}
\ee
\noindent
which is a form ideally suited for computation. 

This form of the Schr\"{o}dinger equation can be much more simply
expressed as 
\be K_{ij}\Psi_j=EM_{ij}\Psi_j
\label{eig1}
\ee 
where the indices $i$ and $j$ range over all nodal wavefunction
degrees of freedom, and the repeated indices indicate a summation.
Details of the assembly of the matrices $K$ and $M$ are given in Appendix B.

\subsection*{B.  Finite Element Solution}

\subsubsection*{\bf{\it{1.  Energies and wavefunctions}}}

The finite element expression of the Schr\"{o}dinger equation
(\ref{eig1}) is in the form of a generalized eigenvalue equation.  The
problem has $n\alpha$ solutions, where $n$ is the total number of
nodes in the mesh and $\alpha$ is the number of subbands in the
quantum mechanical basis.  The solutions consist of energies $E$ and
wavefunctions $\Psi$, of which the lowest energy states are the most
accurate.  

Of the $n\alpha$ eigenstates, some states can be found for which the
wavefunction $\Psi(\ul{r})$ or the probability density
$|\Psi(\ul{r})|^2$ is confined to the active region of the device.
Examples of eigenstates confined to the quantum well layers of the
quantum dot are shown in Figure~6.  The lowest energy
states of these confined eigenstates are the most relevant to
transport.  For the example of the resonant tunneling diode,
sequential tunneling through the double barriers is possible only when
the tunneling carriers have energies equal to the energies of the
confined states in the quantum well.  Thus, over a range of applied
biases, the excited carriers can access the confined states and induce
a tunneling current only at certain resonances corresponding to the
spectrum of eigenvalues given by the finite element solution.

The energy and wavefunction solutions reflect the effects of strain,
composition, and effective mass on the carriers.  The valence band
offset in adjacent layers imposes a large relative potential on the
charge carriers which results in confinement to the quantum well
region of the device.  The strain induced potential is considerably
smaller than the valence band potential, but it also shifts the
wavefunctions spatially and energetically.  Confined states
corresponding to valence subbands with higher effective masses occur
at lower energies.

\subsubsection*{\bf{\it{2.  Density of states}}}

The density, with respect to energy, of states confined to the active
region of the device can be obtained directly from the spectrum of
eigenstates given in the finite element solution.  This density of
confined states is a real space measure of the electronic properties
of the device.  Effects due to strain, composition, size, and device
characteristics can be seen in the density of confined states.  For
the example of the resonant tunneling diode, the density of confined
states can be used to examine strain effects on the resonant tunneling
spectrum.  A larger density of confined states with a given energy
increases the probability of resonant tunneling by carriers with the
same energy.  A high probability of tunneling at a given energy is measured
experimentally as a tunneling current peak.  Thus, the density of
confined states can be used to make contact with experimental data.

A Gaussian broadening technique can be used to calculate a continuous
density of confined states function $\rho(E)$.  The energy $E_i$ of
each confined state is broadened by a narrow, normalized Gaussian
distribution, and a summation over all $n$ states gives $\rho(E)$ as
\be \rho(E)=\sum_{i=1}^n{1 \over
{2a\sqrt{\pi}}}\exp{(-(E-E_i)^2/4a^2)} \ee where $a$ is a free
parameter that controls the width of the Gaussian distributions and
thus the smoothness of the density of states $\rho(E)$.  The parameter
$a$ is chosen to bring out the general features of $\rho(E)$; the
broadening of each state is larger than the typical separation of
individual eigenstates, but narrow enough to bring out features of the
density of states that are due to small groups of related eigenstates.
Typical values of $a$ are on the order of 1 meV.

\section*{IV.  RESULTS AND COMPARISON TO EXPERIMENTS}

\subsection*{A.  The Quantum Wire}

\subsubsection*{\bf{\it{1.  Physical system}}}

The quantum wire considered here is a long, row-shaped, layered
structure of fixed total height $h$.  The geometry of the structure
and the thickness and composition of each layer, based on the
experimental work of Lukey \etal$^{4}$, are shown in Figure~7.  The
middle layers of the device are considered to be the active region,
and include the quantum well layer (Si$_{0.78}$Ge$_{0.22}$) and the
two undoped barrier layers (Si).  Surrounding the active region are
strained (Si$_{0.78}$Ge$_{0.22}$) emitter and collector regions; the
thick outermost layers of the device are Si.  A range of widths $w$
are considered in order to model size-dependent strain effects and to
compare results with experimental data.

The device operates by the sequential tunneling of charge carriers from
the doped region above the barrier layer, into the quantum well layer,
and then through the lower barrier layer.  Resonant tunneling
spectroscopy is done experimentally by applying a bias across the
device and measuring the current of tunneling carriers that is
induced.  A resonant tunneling spectrum can be compiled by measuring
the induced currents associated with a range of applied biases.  The
experimental result is a resonant tunneling I(V) curve. 

\subsubsection*{\bf{\it{2.  Strain field}}}

The strain and displacement fields for this geometry are two
dimensional since the constraining effect of the material in the
direction along the long axis of the wire imposes a state of plane
strain.  The mesh is more refined in the active region of the
structure and near the traction free lateral surfaces, where the
deformation is expected to be more nonuniform.  The $\epsilon_{22}$
component of strain (extensional strain in the $x_2$ direction) is
shown in Figure~3 in and near the active region of the device, plotted
on a deformed mesh that shows the actual displacements, magnified for
clarity.  In Figure~4 the $\epsilon_{11}$ component of strain along
the centerline of the quantum well layer is plotted for structures
with three different widths.  The important features are that the
strain is a tensor valued function and that the relaxing effect of the
free surface and the multilayered composition of the structure lead to
nonuniform strain components.

\subsubsection*{\bf{\it{3.  Results of the quantum mechanical calculation}}}

Solution of the quantum mechanical problem gives the energies and
wavefunctions of states confined to the quantum well layer in the
wire.  From this spectrum of states, a density of confined states is
calculated.  The density of confined states for a narrow wire
(w=250nm) is shown in Figure~8, representative eigenstates are shown
in Figures 9 and 10, and the density of confined states for wide and
narrow wires are compared in Figure~11.

The results of the calculations are consistent with some experimental
observations of Lukey \etal.  First, the strain separates the resonant
current peaks associated with the heavy hole and light hole subbands,
as shown in Figure~8.  The calculation gives a peak separation of
about 20 meV, which would correspond to a bias shift of roughly 50 mV.
The experimentally measured separation is approximately 90 mV.
However, it is important to note that the calculation shown is for a
wire aligned along a (100) crystalline axis.  The experiments of Lukey
\etal measured wires aligned along a (110) axis, which would exhibit
more sensitivity to strain in the electronic properties due to the
form of the deformation potential tensor $D_{ij}^{\alpha\beta}$, and
thus presumably a wider strain induced peak separation.  The second
characteristic evident in Figure~8 is the presence of additional fine
features in the density of states.  Fine structure is also observed
experimentally in the w=250nm device.  An examination of the states
present over a range of energies shows that the fine features in the
density of states are due to groups of similar states separated in
energy by the influence of nonuniform strain, as shown in Figures
9 and 10.

The size dependence of the strain effect is demonstrated in the
densities of confined states for wires with widths of 250nm and 900nm
in Figure~11.  The main feature is the increase in the energy
separation of the heavy hole and light hole peaks as the wire width
increases.  This is due to the reduced effect of free surface strain
relaxation in larger devices, where the average strain values approach
the bulk film mismatch strains.  In the smallest devices, the strain
is relaxed over a significant portion of the volume, so the average
strain is smaller and the energy separation between the heavy hole and
light hole peaks is smaller.

\subsection*{B.  The Quantum Dot}

\subsubsection*{\bf{\it{1.  Physical system}}}

The quantum dot considered here is cylindrical in shape and of fixed
height.  The features of this structure are based on the experimental
work of Zaslavsky \etal$^1$ and Aky\"{u}z \etal$^{2}$ and are shown in
Figure~12 for a representative calculation.  The three middle layers,
which consist of the quantum well (Si$_{0.75}$Ge$_{0.25}$) and the
barriers (Si), make up the active region of the device.  Surrounding
the active region are the emitter and collector regions
(Si$_{0.75}$Ge$_{0.25}$) which have relatively low strain due to the
outermost layers which are graded in composition.  A range of cylinder
diameters $d$ is considered.

The quantum dot device operates on the same resonant tunneling
principle as the quantum wire.  Carriers tunnel sequentially from the
emitter region, through the upper barrier into states in the quantum
well, and then through the lower barrier into the collector
region. The experimental I(V) curve is a measure of the resonant
tunneling spectrum. 

\subsubsection*{\bf{\it{2.  Strain field}}}

The stress, strain, and displacement fields are axisymmetric for this
geometry.  The mesh used to calculate the strain extends from the
center axis of the structure outward, with increasing refinement near
the outer, traction free surface of the device, where the fields are
expected to be more nonuniform.  Figure~13 shows the $\epsilon_{22}$
component of strain (extensional strain in the axial direction) on a
deformed mesh, and Figure~14 shows $\epsilon_{11}$, the extensional
strain along a radial line in the midplane of the quantum well.  As in
the case of the quantum wire, the strain is very nonuniform, and the
extent of the nonuniformity increases in smaller structures.  This is
due to the more significant effect of free surface proximity.

\subsubsection*{\bf{\it{3.  Results of the quantum mechanical calculation}}}

For the three-dimensional quantum dot, a reduced quantum mechanical
basis is adopted in order to limit the total number of degrees of
freedom in the calculation.  To model only the lowest energy heavy
hole states, it is sufficient to consider a one-dimensional quantum
mechanical basis where only the ${\hhp}$ band is examined, but it is still
possible to consider an anisotropic effective mass.

Two significant results compare favorably to the experiments of
Zaslavksy \etal$^{1}$ and Aky\"{u}z \etal$^{2}$.  First, as seen in
Figure~15, there is a size dependence of the strain effect on the
resonant energies that is similar to the effect in the quantum wires.
In larger diameter devices, the average strain is higher because free
surface relaxation is less significant, so the peak is shifted further
from an idealized case without strain.

Second, in the smallest devices, there are additional fine features in
the density of states that are shown in Figure~16 to be the result of
the influence of nonuniform strain.  In agreement with the suggestion
of Aky\"{u}z \etal$^2$, it is evident that the relaxed strain near the
lateral surface leads to low energy ring-like heavy hole states.  The
strain induced energy shifting of groups of states produces features
in the density of confined states that are consistent with the I(V)
curves for devices of the same size.  Figure~17 shows a calculated
density of confined states plot and a measured I(V) curve for a
d=250nm quantum dot with 10meV on the energy axis equal to 25 mV on
the bias axis.  Many of the features of the I(V) curve are predicted
qualitatively in the density of confined states curve, including fine
structure below and above the main heavy hole resonance energy.
However, the density of confined states is not equivalent to the
resonant tunneling current; the calculation does not consider some
important physical effects, most notably the roughly linear background
current in the I(V) relationship.

\section*{V.  CONCLUSIONS}

A finite element technique is presented here which allows for the
calculation of strain effects on the electronic and transport
properties of strained quantum wires and dots.  The approach is
similar to some recent work as it is based on a simplified quantum
mechanical model$^{5-8}$, but the flexibility of meshing and the low
computational cost of the finite element method offer easy access to
results which can be compared to experimental data.

The technique is used to examine mismatch strain effects in quantum
wires and quantum dots that operate on a simple single carrier
sequential tunneling effect.  Strain effects are shown to explain
several reported trends in experimental data.  In particular, two
features of the mismatch strain in the devices have strong effects on
the calculated electronic and transport properties.  First, the
average effect of the strain is to separate the resonant energy peaks
associated with the individual valence subbands in the material.  In
larger devices, the strain is less relaxed by the free surfaces, so
the HH-LH subband energy separation is larger.  Second, strain
nonuniformity in the devices is responsible for fine structure in the
resonant tunneling current peaks.  This effect is the source of low
energy ring-like states that are found in the small cylindrical
quantum dots, and the edge states found in quantum wires over a range
of sizes.

The main weaknesses in the method are in the simplified quantum
mechanical model.  The real-space calculation is based on a $k$-space
material model that is accurate near $k=0$.  The resonant tunneling
model assumes ballistic transport of a single charge carrier, and
contact is made with experiments in only an approximate way.  A linear
elastic constitutive model provides a good approximation for the
strain, although the approach has limitations for such small, highly
strained structures$^{25}$.  Finally, the strain effect is treated as a
linear perturbation to a perfect crystal Hamiltonian, so the fully
coupled nature of strain and electronic properties through the
chemical bonding is not considered.  However, the technique shown here
is a promising, computationally inexpensive way to determine strain
effects on electronic properties in semiconductors.  The means to
overcome the noted shortcomings are under development.

\section*{ACKNOWLEDGEMENTS}

One of the authors (HTJ) would like to acknowledge helpful discussions
with V. B. Shenoy and R. Phillips regarding the application of the
finite element method to quantum mechanics.  The authors are also
grateful for the cooperation of J. Caro in providing access to
experimental data.  The research support of the Office of Naval
Research, Contract N00014-95-1-0239, and the MRSEC Program of the
National Science Foundation, under Award DMR-9632524, is gratefully
acknowledged.

\newpage

\section*{REFERENCES}

\noindent
\hangindent 14pt{
$^1$  A. Zaslavsky, K. R. Milkove, Y. H. Lee, B. Ferland, and
T. O. Sedgewick, Appl. Phys. Lett. {\bf 67}, 3921 (1995).}

\noindent
\hangindent 14pt{
$^2$  C. D. Aky\"{u}z, A. Zaslavsky, L. B. Freund, D. A. Syphers, and
T. O. Sedgewick, Appl. Phys. Lett. {\bf 72}, 1739 (1998).}

\noindent
\hangindent 14pt{
$^3$  P. Gassot, U. Gennser, D. M. Symons, A. Zaslavsky,
and D. A. Gruztmacher, J. C. Portal, submitted to Phys. Rev. B (1997).}

\noindent
\hangindent 14pt{
$^4$  P. W. Lukey, J. Caro, T. Zijlstra, E. van der Drift, and
S. Radelaar, Phys. Rev. B {\bf 57}, 7132 (1998).}

\noindent
\hangindent 14pt{
$^5$  C. Pryor, M-E. Pistol, L. Samuelson, Phys. Rev. B {\bf 56}, 10404 (1997).}

\noindent
\hangindent 14pt{
$^6$  C. Pryor, Phys. Rev. B {\bf 57}, 7190 (1998).}

\noindent
\hangindent 14pt{
$^7$  C. Pryor, http://xxx.lanl.gov/abs/cond-mat/9801225.}

\noindent
\hangindent 14pt{
$^8$  A. J. Williamson, A. Zunger, A. Canning, http://xxx.lanl.gov/abs/cond-mat/9801191.}

\noindent
\hangindent 14pt{
$^9$  A. Zunger, MRS Bull. {\bf 23} 15, 1998.}

\noindent
\hangindent 14pt{
$^{10}$  T. J. Gosling and  L. B. Freund, Acta mater. {\bf 44}, 1 (1996).}

\noindent
\hangindent 14pt{
$^{11}$  ABAQUS, Version 5.5, Hibbitt, Karlsson \& Sorensen, Inc.,
Pawtucket, RI 02860, U. S. A. (1995).}

\noindent
\hangindent 14pt{
$^{12}$  J. Linderberg, Comp. Phys. Reports {\bf 6}, 209 (1987).}

\noindent
\hangindent 14pt{
$^{13}$  K. Nakamura, A. Shimizu, M. Koshiba, and K. Hayata, IEEE J. Quantum
Electron. {\bf 25}, 889 (1989).}

\noindent
\hangindent 14pt{
$^{14}$  A. Zhao, S. R. Cvetkovic, and Z. A. Yang, Opt. Quant. Electron. {\bf 25}, 845 (1993).}

\noindent
\hangindent 14pt{
$^{15}$  T. L. Li and K. J. Kuhn, J. Comp. Phys. {\bf 115}, 288 (1994).}

\noindent
\hangindent 14pt{
$^{16}$  Chen, Computers Math. Applic. {\bf 31}, 17 (1996).}

\noindent
\hangindent 14pt{
$^{17}$  K. Kujima, K. Mitsunaga, and K. Kyuma, Appl. Phys. Lett. {\bf
55}, 882, (1989)}

\noindent
\hangindent 14pt{
$^{18}$  D. J. Kirkner, C. S. Lent, and S. Sivaprakasm,
I. J. Num. Meth. Eng. {\bf 29}, 1527, (1990)}

\noindent
\hangindent 14pt{
$^{19}$  Z. Wu and P. P. Ruden, J. Appl. Phys. {\bf 74}, 6234 (1993).}

\noindent
\hangindent 14pt{
$^{20}$  Y. Wang, J. Wang, and H. Guo,  Phys. Rev. B {\bf 49}, 1928 (1994).}

\noindent
\hangindent 14pt{
$^{21}$  T. Inoshita and H. Sakaki, J. Appl. Phys. {\bf 79}, 269 (1996).}

\noindent
\hangindent 14pt{
$^{22}$  J. C. Yi and N. Dagli,  IEEE J. Quantum Electron. {\bf 31}, 208 (1995).}

\noindent
\hangindent 14pt{
$^{23}$  T. Inoshita and H. Sakaki, J. Appl. Phys. {\bf 79}, 269 (1996).}

\noindent
\hangindent 14pt{
$^{24}$  E. Tsuchida and M. Tsukada, Phys. Rev. B {\bf 52}, 5573 (1995).}

\noindent
\hangindent 14pt{
$^{25}$  J. Singh, {\it Physics of Semiconductors and Their
Heterostructures}, (McGraw-Hill, New York, 1993), p. 228.}

\noindent
\hangindent 14pt{
$^{26}$  C. Pryor, J. Kim, L. W. Wang, A. Williamson and A. Zunger,
J. Appl Phys. {\bf 83}, 5 1998.}

\newpage

\section*{Appendix A}

The strain induced potential $V_\epsilon^{\alpha\beta}(\ul{r})$ is
given by
$$
V_\epsilon^{\alpha\beta}(\ul{r})=
\bordermatrix{& {\hhp} & {\hhm} & {\lhp} & {\lhm} \cr
{\hhp} & D_{ij}^{11}(\ul{r})\epsilon_{ij}(\ul{r}) & D_{ij}^{12}(\ul{r})\epsilon_{ij}(\ul{r}) & D_{ij}^{13}(\ul{r})\epsilon_{ij}(\ul{r}) & D_{ij}^{14}(\ul{r})\epsilon_{ij}(\ul{r}) \cr
{\hhm} & D_{ij}^{21}(\ul{r})\epsilon_{ij}(\ul{r}) & D_{ij}^{22}(\ul{r})\epsilon_{ij}(\ul{r}) & D_{ij}^{23}(\ul{r})\epsilon_{ij}(\ul{r}) & D_{ij}^{24}(\ul{r})\epsilon_{ij}(\ul{r}) \cr
{\lhp} & D_{ij}^{31}(\ul{r})\epsilon_{ij}(\ul{r}) & D_{ij}^{32}(\ul{r})\epsilon_{ij}(\ul{r}) & D_{ij}^{33}(\ul{r})\epsilon_{ij}(\ul{r}) & D_{ij}^{34}(\ul{r})\epsilon_{ij}(\ul{r}) \cr
{\lhm} & D_{ij}^{41}(\ul{r})\epsilon_{ij}(\ul{r}) & D_{ij}^{42}(\ul{r})\epsilon_{ij}(\ul{r}) & D_{ij}^{43}(\ul{r})\epsilon_{ij}(\ul{r}) & D_{ij}^{44}(\ul{r})\epsilon_{ij}(\ul{r}) \cr
} 
\eqno({A1})
$$
where each component $D_{ij}^{\alpha\beta}(\ul{r})$ of the matrix for
fixed $\alpha\beta$ forms a scalar product with the
strain tensor $\epsilon_{ij}(\ul{r})$ through summation over $i$ and $j$.  
And similarly, the {\bf k} $\cdot$ {\bf p} Hamiltonian given by Luttinger and Kohn
takes the form
$$
H_{k \cdot p}^{\alpha\beta}(\ul{r})=
\bordermatrix{& {\hhp} & {\hhm} & {\lhp} & {\lhm} \cr
{\hhp} & L_{ij}^{11}(\ul{r})\nabla^2_{ij} & L_{ij}^{12}(\ul{r})\nabla^2_{ij} & L_{ij}^{13}(\ul{r})\nabla^2_{ij} & L_{ij}^{14}(\ul{r})\nabla^2_{ij} \cr
{\hhm} & L_{ij}^{21}(\ul{r})\nabla^2_{ij} & L_{ij}^{22}(\ul{r})\nabla^2_{ij} & L_{ij}^{23}(\ul{r})\nabla^2_{ij} & L_{ij}^{24}(\ul{r})\nabla^2_{ij} \cr
{\lhp} & L_{ij}^{31}(\ul{r})\nabla^2_{ij} & L_{ij}^{32}(\ul{r})\nabla^2_{ij} & L_{ij}^{33}(\ul{r})\nabla^2_{ij} & L_{ij}^{34}(\ul{r})\nabla^2_{ij} \cr
{\lhm} & L_{ij}^{41}(\ul{r})\nabla^2_{ij} & L_{ij}^{42}(\ul{r})\nabla^2_{ij} & L_{ij}^{43}(\ul{r})\nabla^2_{ij} & L_{ij}^{44}(\ul{r})\nabla^2_{ij} \cr
}
\eqno({A2})
$$
where the each of the matrix components $L_{ij}^{\alpha\beta}(\ul{r})$
for fixed $\alpha\beta$ form a scalar product with the operator
$\nabla^2_{ij}$.The components $D_{ij}^{\alpha\beta}(\ul{r})$ and
$L_{ij}^{\alpha\beta}(\ul{r})$ have very similar form.  The
deformation potential components $D_{ij}^{\alpha\beta}(\ul{r})$ are
$$
D_{ij}^{11}(\ul{r})=D_{ij}^{22}(\ul{r})=
\left[
\matrix{a+{\frac b 2} & 0 & 0 \cr
	0 & a+{\frac b 2} & 0 \cr
	0 & 0 & a-b \cr}
\right]\nonumber\\
$$
$$
D_{ij}^{33}(\ul{r})=D_{ij}^{44}(\ul{r})=
\left[
\matrix{a-{\frac b 2} & 0 & 0 \cr
	0 & a-{\frac b 2} & 0 \cr
	0 & 0 & a+b \cr}
\right]\nonumber\\
$$
$$
D_{ij}^{13}(\ul{r})=D_{ij}^{31^*}(\ul{r})=-{D_{ij}^{24}}^*(\ul{r})=-D_{ij}^{42}(\ul{r})=
\left[
\matrix{0 & 0 & -i{\frac d 2} \cr
	0 & 0 & -{\frac d 2} \cr
	-i{\frac d 2} & -{\frac d 2} & 0 \cr}
\right]\nonumber\\
$$
$$
D_{ij}^{14}(\ul{r})={D_{ij}^{23}}^*(\ul{r})=D_{ij}^{32}(\ul{r})={D_{ij}^{41}}^*(\ul{r})=
\left[
\matrix{{\frac {\sqrt{3}} 2}b & -i{\frac d 2} & 0 \cr
	-i{\frac d 2} & -{\frac {\sqrt{3}} 2}b & 0 \cr
	0 & 0 & 0 \cr}
\right]\nonumber\\
$$
$$
D_{ij}^{12}(\ul{r})=D_{ij}^{21}(\ul{r})=D_{ij}^{34}(\ul{r})=D_{ij}^{43}(\ul{r})=
\left[
\matrix{0 & 0 & 0 \cr
	0 & 0 & 0 \cr
	0 & 0 & 0 \cr}
\right]
\eqno({A3})
$$
\noindent
The Hamiltonian components $L_{ij}^{\alpha\beta}(\ul{r})$ can be
obtained by making the substitutions ${\frac
{\hbar^2}{2m_0}}\gamma_1{\leftrightarrow}a$, ${\frac {\hbar^2}{m_0}}\gamma_2{\leftrightarrow}b$,
${\frac {\sqrt{3}\hbar^2}{m_0}}\gamma_3{\leftrightarrow}d$ into the expressions for
the components $D_{ij}^{\alpha\beta}(\ul{r})$, where $\gamma_1$,
$\gamma_2$, $\gamma_3$ are the Luttinger-Kohn parameters.  Values for
the deformation potential constants and the Luttinger-Kohn parameters
for Si and Ge are given below.  Values for alloys of Si and Ge are
interpolated from values for the bulk materials.
$$
\arraycolsep=6pt
\begin{array}{|l||*{6}{l|}} \hline
 & a (eV) & b (eV) & d (eV) & \gamma_1 & \gamma_2 & \gamma_3 \\ \hline
Si & 2.1 & -1.5 & -3.4 & 4.29 & 0.34 & 1.45 \\ \hline
Ge & 2.0 & -2.2 & -4.4 & 13.4 & 4.24 & 5.59 \\ \hline
\end{array}
$$

\section*{Appendix B}

To obtain the finite element form of the Schr\"{o}dinger equation, the
physical region is divided into elements, which are taken here to be
4-noded quadrilaterals for the two dimensional quantum wire and
8-noded bricks for the three dimensional quantum dot.  The wavefunction
$\Psi^\alpha$, wavefunction gradient $\nabla\Psi^\alpha$, and
potential $V^{\alpha\beta}$ are expressed in terms of discretized
values at the nodes, and values within the elements are determined by
linear interpolation using linear shape functions $N(\ul{r})$.  A group
of quadrilateral elements and the linear shape function for an
associated node is shown in Figure~18.

The form of the Schr\"{o}dinger equation to be solved is
$$
L^{\alpha \beta}(\ul{r})\nabla^2\Psi^{\beta}(\ul{r})+V^{\alpha
\beta}(\ul{r})\Psi^{\beta}(\ul{r})=E\Psi^{\alpha}(\ul{r})
\eqno({B1})
$$ 
\noindent
The weak form of the equation is obtained by multiplying by
$\Psi(\ul{r})$ and integrating over the volume of the body.  The first
term is integrated by parts, and the functional corresponding to the
weak form is given by
$$
\Pi(\Psi^{\alpha})=\int_R \nabla \Psi^{\alpha}
L^{\alpha\beta} \nabla \Psi^{\beta} dR + \int_R
\Psi^{\alpha}V^{\alpha\beta}\Psi^{\beta} dR - E\int_R
\Psi^{\alpha}\Psi^{\beta} dR
\eqno({B2})
$$
\noindent
The spatially varying fields are then discretized using the shape
functions $N(\ul{r})$ to get
$$
\Pi(\Psi_A^{\alpha})=\sum_A \sum_B \left[ \int_R \Psi_A^{\alpha}\nabla N_A 
L^{\alpha\beta} \Psi_B^{\beta} \nabla N_B dR + \int_R
\Psi_A^{\alpha} N_A \sum_C V_C ^{\alpha\beta} N_C \Psi_B^{\beta} N_B dR - E\int_R
\Psi_A^{\alpha}\Psi_B^{\beta}N_A N_B\right ] dR
\eqno({B3})
$$
\noindent
The total variation of the functional $\Pi(\Psi_A^{\alpha})$ is then
minimized with respect to the nodal values of the wavefunction
$\Psi_B^{\beta}$ so that
$$
{\frac {d\Pi(\Psi_A^{\alpha})} {d\Psi_B^{\beta}}}=0
\eqno({B4})
$$
\noindent
thus
$$
\Psi_A^{\alpha} \sum_A \sum_B \left[ \int_R \nabla N_A 
L^{\alpha\beta} \nabla N_B dR + \int_R
N_A \sum_C V_C ^{\alpha\beta} N_C N_B dR -E\int_R
N_A N_B dR \right ]=0
\eqno({B5})
$$
\noindent
Replacing integrals over the region $R$ with integration over
individual element volumes ($\Omega^e$) and a summation over all
elements, the final finite element form of the equation becomes
$$
\sum^{elems} \left [ \sum_A \sum_B \int_{\Omega^e} \left(\nabla
N_A L^{\alpha\beta}\nabla N_B + N_A N_B \sum_C V_C^{\alpha\beta} N_C \right)d\Omega
\right ] \Psi_B^{\beta} = E \sum^{elems} \left [
\sum_A \sum_B \int_{\Omega^e}N_A N_B d\Omega \right ] \Psi_B^{\alpha}
\eqno({B6})
\label{appfem}
$$
The contribution of a single element to the left hand side of the
equation is given by the element stiffness matrix.  The integration
over the element is done numerically at a set of quadrature points.
The construction of an element stiffness matrix for the
case of two spatial dimensions and a four subband quantum mechanical
basis is as follows:
$$
k_{\alpha\beta}^e=\int_{\Omega^e} \left(\nabla
N_A L^{\alpha\beta}\nabla N_B + N_A N_B \sum_C V_C^{\alpha\beta} N_C
\right)d\Omega \nonumber \\
$$
$$
= \sum_{l=1}^{int.pts.} [ \underbrace{\underbrace {\nabla
N_A}_{16 \times 8} \underbrace{L^{\alpha\beta}}_{8\times 8}
\underbrace{\nabla N_B}_{8 \times 16}}_{16 \times
16}+\underbrace{\underbrace{N_A}_{16\times 4}
 \underbrace{\sum^{nodes} \left( V_C^{\alpha\beta} N_C \right)}_{4 \times 4}
\underbrace{N_B}_{4\times 16}}_{16 \times 16} ]_l
\eqno({B7})
$$
\noindent
where the tensors $D_{ij}^{\alpha\beta}$ and $L_{ij}^{\alpha\beta}$
given in Appendix 1 reduce to $2 \times 2$ matrices, and the shape
function matrix is given by

$$
\def\tempb{\multicolumn{1}{|c}{0}}
\def\tempa{\multicolumn{1}{c|}{0}}
N_A=\left[\begin{array}{cccccccccccccccc}
N_1 & N_2 & N_3 & N_4 & \tempb & \ldots & \ldots & 0 & \ldots &
\ldots & 0 & \ldots & \ldots & 0 & \ldots & \ldots \\ \cline
{1-8}
0 & \ldots & \ldots & \tempa & N_1 & N_2 & N_3 & N_4 & \tempb & \ldots & \ldots & 0 & \ldots &
\ldots & 0 & \ldots  \\ \cline
{5-12}
\ldots & 0 & \ldots & \ldots & 0 & \ldots & \ldots & \tempa & N_1 & N_2 & N_3 & N_4 & \tempb & \ldots & \ldots & 0  \\ \cline
{9-16}
\ldots & \ldots & 0 & \ldots & \ldots & 0 & \ldots & \ldots & 0 & \ldots & \ldots & \tempa & N_1
& N_2 & N_3 & N_4 \\
\end{array} \right] \nonumber \\
\eqno({B8})
$$
\noindent
and the shape function derivative matrix $\nabla N_A$ follows in the
same form.

The contribution of a single element to the right hand side of
equation (B6) is referred to as the element mass matrix and
is constructed in a similar way as $k_{\alpha\beta}^e$.  The final
finite element matrix form of the Schr\"{o}dinger equation, given by
$$
K_{ij}\Psi_j=EM_{ij}\Psi_j
\eqno({B9})
$$
\noindent
is constructed by assembling the element stiffness matrices and element
mass matrices into global element and mass matrices, $K_{ij}$ and $M_{ij}$,
for the entire device.

\newpage

\section*{List of Captions}
\begin{enumerate}
\item{Schematics of the quantum wire and quantum dot.  The
row-shaped quantum wire (Lukey \etal) has width w and
extends a distance much larger than w in the (010) direction.  The
cylindrical quantum dot (Aky\"{u}z \etal) has diameter d.}
\item{Finite element mesh used to calculate strain
in the quantum wire.  Since the strain is symmetric laterally and
vertically about the center axes of the device, it is possible to
model one quarter of the cross section of the structure only.
A portion of the highly refined mesh near the edge of the
active region is shown.}
\item{Direct vertical component of strain in the upper half of the
quantum wire, from the center to the edge.  Displacements of the free
outer surface of the device can be seen at the right edge of the plot.
The layers, from the bottom, represent the well and barrier layers
(highly nonuniformly strained) and the emitter and topmost Si layers.}
\item{Direct lateral component of strain in the center of the well
layer of the quantum wire.  The strain is uniform near the center of
the wire (left) but highly nonuniform near the edge (right).  In
larger wires the strain is less relaxed and less nonuniform.}
\item{The total potential $V(r)$ for the heavy hole band in the
strained quantum dot.  The potential in the barrier layers is high
relative to the potential in the quantum well layer; the potential is
axisymmetric, and radially nonuniform throughout the device.}
\item{Representative probability densities in a d=50nm quantum
dot.  On the left is a low energy state with six-fold angular
quantization localized in a ring-like region near the outer edge of
the active region.  On the right is a higher
energy state, localized in the center of the active region, with two fold angular quantization and two fold quantization in the z direction.}
\item{Schematic of the quantum wire geometry and composition in the
active region.  The 59$\AA$ Si barriers surround the 33$\AA$
Si$_{0.78}$Ge$_{0.22}$ quantum well layer.  An applied bias V induces
a tunneling current in the z direction.}
\item{Density of confined states in the strained quantum wire of
width 250nm.  The two large peaks in the dashed curve show the
calculated heavy hole and light hole resonances without considering
strain effects.  Strain causes the resonances to separate in energy
and induces fine structure in the density of confined states.  States
at points A (edge state) and B (light hole state) are shown in
Figures 9 and 10.}
\item{Probability density for a lower energy confined state.  The
four cross sections of the active region show the probability density
associated with each of the four valence subbands for a given
eigenstate.  The wavefunction in this eigenstate is of mixed type, and
localized near the edge of the structure due to the free
surface strain relaxation.}
\item{Probability density for a higher energy confined state.  The
predominantly $\lhp$ type wavefunction is localized in the center of the
quantum well, but shows some effects of the nonuniform potential near
the lateral surfaces and strong mixing with the $\hhm$ subband.}
\item{Densities of confined states for two quantum wires of
different widths.  The higher average strain in the wider device
results in a larger energy separation between heavy hole and light
hole resonance peaks.  The heavy hole peak is shown to be shifted by
the strain more than the light hole peak.}
\item{Geometry and composition of the quantum dot device (Aky\"{u}z
\etal).  The 45$\AA$ Si barriers surround the 35$\AA$
Si$_{0.75}$Ge$_{0.25}$ well.  The total height is 80nm, and the
diameter d varies.  An applied bias V induces a tunneling current in
the z direction.}
\item{Vertical component of strain in the upper half of the quantum
dot, from the center to the edge.  The plot is deformed to show the
displacement in the structure, and the strain is highly
nonuniform, particularly in the active region of the device.  The two
layers at the bottom are the quantum well and top barrier layers.}
\item{Direct lateral component of strain in the well layer of the
quantum dot, from the center to the edge.  The strain is more relaxed
and more nonuniform in the smaller structures.  The strain profile is
very similar to the profile in the quantum wire.}
\item{Densities of heavy hole confined states for a range of
quantum dot diameters.  The variation in average strain levels results
in the shifting of the peaks for dots of different sizes.  The energy
shift corresponds to the bias shift measured by Zaslavsky \etal.}
\item{The effect of strain induced lateral confinement on the
density of confined states.  The features of the density of confined
states are due to groups of eigenstates with similar lateral quantization.}
\item{Fine structure in the density of confined states in the
d=250nm quantum dot (Aky\"{u}z \etal).  The calculated density of
confined states is consistent with the experimental current-voltage
curve.  The density of confined states calculation does not account
for the roughly linear background current that is observed experimentally.}
\item{Elements, nodes, and a representative shape function in the
two dimensional finite element formulation.}
\end{enumerate}

\end{document}